\documentclass[twocolumn,superscriptaddress,longbibliography]{revtex4-1}
\usepackage{color}
\usepackage[colorlinks=true,urlcolor=blue,citecolor=blue]{hyperref}
\usepackage{soul,ulem}
\usepackage{graphicx}
\usepackage{amssymb}
\usepackage{amsmath}
\usepackage{subfigure}
\usepackage{setspace}
\usepackage{hyperref}
\usepackage{url}
\usepackage{epstopdf}
\usepackage{mathrsfs}
\usepackage[mathscr]{euscript}
\usepackage{gensymb}
\usepackage[hyphenbreaks]{breakurl}

\sethlcolor{yellow}

\graphicspath{{figures/}}

\makeatletter
\newcommand*\bigcdot{\mathpalette\bigcdot@{.5}}
\newcommand*\bigcdot@[2]{\mathbin{\vcenter{\hbox{\scalebox{#2}{$\m@th#1\bullet$}}}}}
\makeatother

\begin{document}
	\title{Noise-Tolerant Detection of $Z_N$ Topological Orders in Quantum Many-Body States}
	
	\author{Xi Chen}
	\affiliation{School of Physical Sciences, University of Chinese Academy of Sciences, P. O. Box 4588, Beijing 100049, China}
	\affiliation{Kavli Institute for Theoretical Sciences, and CAS Center of Excellence in Topological Quantum Computation , University of Chinese Academy of Sciences, Beijing 100190, China}
	
	\author{Shi-Ju Ran}
	\email[Corresponding author. Email: ] {sjran@cnu.edu.cn}
	\affiliation{Department of Physics, Capital Normal University, Beijing 100048, China}
	\affiliation{ICFO-Institut de Ciencies Fotoniques, The Barcelona Institute of Science and Technology, 08860 Castelldefels (Barcelona), Spain}
	
	\author{Shuo Yang}
	\affiliation{State Key Laboratory of Low-Dimensional Quantum Physics and Department of Physics, Tsinghua University, Beijing 100084, China}
	
	\author{Maciej Lewenstein}
	\affiliation{ICFO-Institut de Ciencies Fotoniques, The Barcelona Institute of Science and Technology, 08860 Castelldefels (Barcelona), Spain}
	\affiliation{ICREA, Passeig Lluis Companys 23, 08010 Barcelona, Spain}
	
	\author{Gang Su}
	\email[Corresponding author. Email: ] {gsu@ucas.ac.cn}
	\affiliation{School of Physical Sciences, University of Chinese Academy of Sciences, P. O. Box 4588, Beijing 100049, China}
	\affiliation{Kavli Institute for Theoretical Sciences, and CAS Center of Excellence in Topological Quantum Computation , University of Chinese Academy of Sciences, Beijing 100190, China}
	
	\begin{abstract}
		Topologically ordered states are fundamentally important in theoretical physics, which are also suggested as promising candidates to build fault-tolerant quantum devices. However, it is still elusive how topological orders can be affected or detected under noises. In this work, we find a quantity, termed as the ring degeneracy $\mathcal{D}$, which is robust under pure noise to detect both trivial and intrinsic topological orders. The ring degeneracy is defined as the degeneracy of the solutions of the self-consistent equations that encode the contraction of the corresponding tensor network(TN). For the $Z_N$ orders, we find that the ring degeneracy satisfies a simple relation $\mathcal{D} = (N + 1)/2 + d$, with $d = 0$ for odd $N$ and $d = 1/2$ for even $N$. Simulations on several non-trivial states (two-dimensional Ising model, $Z_N$ topological states, and resonating valence bond states) show that the ring degeneracy can tolerate noises up to a strength associated to the gap of the TN boundary theory.
	\end{abstract}
	\maketitle
	
	\section{Introduction}
	Topological states \cite{K. Klitzing80, D. C. Tsui82, X. G. Wen89, X. G. Wen90} are exotic states of matter that cannot be described by conventional order parameters, such as those within the Landau-Ginzburg paradigm. This kind of states have been considered as promising candidates to realize fault-tolerant quantum devices, e.g., quantum computers \cite{A. Y. Kitaev03,R. W. Ogbrun99, M. H. Freedman03} and quantum memories \cite{E. Dennis02, A. Y. Kitaev03}. Taking the Kitaev honeycomb model as an example \cite{A. Y. Kitaev03}, the degenerate ground state provides a subspace that can store the information like the qubits. Since the degenerate states are connected by non-local operations that wind the whole system, that is, they are protected by a large gap, local perturbations will not be able to induce any errors to the stored information as long as the perturbations are smaller than the energy gap \cite{A. Y. Kitaev03, S. Dusuel11, X. Chen11}.
	
	Several methods and signatures have been proposed to detect the topological orders. The most widely applied ones are: i) the topological entanglement entropy (TEE) \cite{A. Kitaev06, M. Levin06}, ii) the topological Renyi entropy \cite{S. Flammia09}, iii) the topological ground-state degeneracy \cite{X. G. Wen90-1, S. Depenbrock12}, and other method like ribbon operators \cite{J. C. Bridgeman16}. For the symmetry protected topological (SPT) states \cite{ZC.Gu09}, the fixed-point tensors from tensor-entanglement-filtering renormalization \cite{ZC.Gu09} are used to characterize the symmetry breaking and SPT phase transitions. 
	
	However, due to highly computational complexity, the investigations on realistic higher-dimensional quantum models are still rare, particularly for those systems that do not admit known analytical solutions. For 1D quantum system, the bipartite entanglement spectrum can be used to characterize topological phase \cite{A. Kitaev06, M. Levin06, Haldane, H. Li08, L. Fidkowski, A. M. Turner}, which has been applied to detect Haldane phase \cite{F. D. M. Haldane83-1, F. D. M. Haldane83-2, ZC.Gu09, F. Pollmann10, F. Pollmann12}. But, for two- and higher-dimensional systems, the applications are sparse \cite{J. M. Kosterlitz73, M. Levin05, Z. C. Gu09}, essentially due to the impressive complexity in calculating the entanglement in higher dimensions. Such difficulty also hinders the applications of TEE and topological Renyi entropy for detecting the topological orders in higher dimensions.
	
	Moreover, it remains elusive how the noises affect the topological states, which is an important issue to the utilization of topological systems to develop novel quantum technologies \cite{A. Y. Kitaev03,R. W. Ogbrun99, E. Dennis02, M. H. Freedman03}. Chen \textit{et al}. showed that the topological Renyi entropy is stable only against the $Z_2$ symmetry preserving variations on the tensors of topological TN states \cite{X.Chen10}. Therefore, it is hard to use topological Renyi entropy to detect the topological orders of the states that are obtained by numerical simulations, where there always exist the numeric noises/errors in the calculations. Besides, it is still interesting to study the states in the symmetry breaking vicinity of a topological state; they may still inherit certain topological properties even when the topological Renyi entropy vanishes.
	
\begin{figure}
		\includegraphics[angle=0,width=1\linewidth]{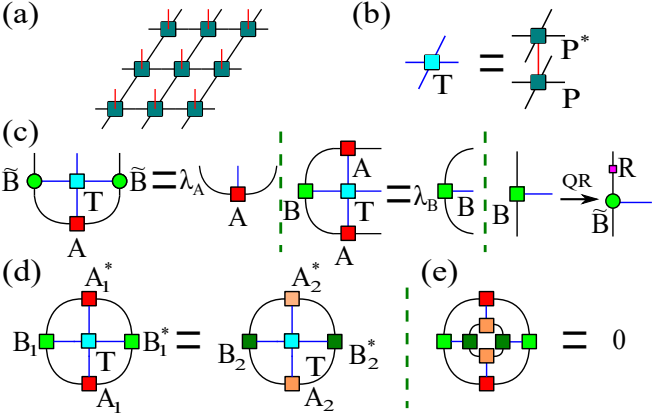}
		\caption{(Color online) (a) A graphic representation of PEPS. (b) From tensor P to inner product tensor T.  (c) Self-consistent eigenvalue equations of TRD. (d) and (e) Two degenerate ring tensors in TRD and the orthogonality of them. }
		\label{fig1}
	\end{figure}

	In this work, we propose a quantity named the ring degeneracy (RD, denoted by $\mathcal{D}$) that robustly detects the symmetries and the topological properties even under a noise that breaks the symmetries of the tensors. RD is defined as the degeneracy of the ring tensor, which is the fixed-point solution of the self-consistent eigenvalue equations constructed from a TN representation of the quantum system [Fig. \ref {fig1} (c)] \cite{S. J. Ran16}. We show that the symmetry of topological states could lead to a degeneracy of the ring tensors. For the 2D statistical Ising model, we show that RD detects the spontaneous symmetry breaking, i.e., with $\mathcal{D}=2$ for the low-temperature symmetry breaking phase and $\mathcal{D}=1$ for the high-temperature disordered phase. For the spin-1 Heisenberg chain in a magnetic field, we have $\mathcal{D} = 2$ in the Haldane phase for $h<0.41$ and $\mathcal{D} = 1$ in the polarized phase for $h>0.41$ \cite{S. R. White93}. For those with $Z_N$ intrinsic topological orders, including the resonating valence bond state on kagom\'e lattice with $Z_2$ topological order \cite{P. W. Anderson73, X. G. Wen91, R. Moessner01, G. Misguich02} and the $Z_N$ string-net states \cite{M. Levin05,M. Levin06, Z. C. Gu09,O. Buerschaper09}, we have $\mathcal{D} = (N + 1)/2 + d$ with $d = 0$ for odd $N$, and $d = 1/2$ for even $N$. RD is a robust quantity even under pure noises. We demonstrate how the noise affects the stability of RD and show that the RD can be reached robustly up to a noise of the same order of magnitude as the gap of the TN boundary theory.
	
	\section{Ring tensor of infinite two-dimensional tensor network}
	An infinite TN state (TNS) (also called the projected entangled pair state) \cite{F. Verstraete06, S. J. Ran17} in 2D system with translation invariance can be written as (Fig. \ref{fig1} (a))
	\begin{equation}
	|\psi\rangle = \sum_{s_1 s_2\cdots} \sum_{\alpha_1 \alpha_2 \cdots} P_{s_1,\alpha_1 \alpha_2 \alpha_3 \alpha_4} P_{s_2,\alpha_4 \alpha_5 \alpha_6 \alpha_7} \cdots|s_1, s_2\cdots\rangle.
	\end{equation}
	The Latin letters $\{s_i\}$ represent the physical indexes that correspond to the physical Hilbert space of the quantum state, and the Greek letters $\{\alpha_j\}$ represent the geometrical indexes that will be contracted. The inner product between the state and its conjugate $\langle \psi | \psi \rangle$ gives a 2D TN, where all physical and geometrical indexes will be contracted. Such a TN is formed by infinite copies of inner product tensor $T_{\eta_1 \eta_2 \eta_3 \eta_4} = \sum_s P_{s,\alpha_1 \alpha_2 \alpha_3 \alpha_4} P^{\ast}_{s,\alpha_1' \alpha_2' \alpha_3' \alpha_4'}$ with $\eta_n = (\alpha_n, \alpha_n')$ (Fig. \ref{fig1} (b)); it is in fact the zero-temperature partition function of the system, and conveys many physical properties of the TNS, such as the correlation length and criticality (e.g., \cite{RanTNcrit}).
	
	Tensor ring decomposition (TRD) \cite{S. J. Ran16, S. J. Ran17} is an efficient way to compute the TN contraction. Unlike the methods based on the tensor renormalization group (see, e.g., Refs. \cite{TRG, TEBD, CTMRG, TRGnew1, TRGnew2, TRGnew3}), TRD ``encodes'' the TN contraction problem to a set of local self-consistent eigenvalue equations. With the spatial inversion symmetries, the solution of TRD contains two tensors dubbed as $A$ and $B$. The eigenvalue equations [see the first two sub-figures of Fig. \ref{fig1} (c)] that $A$ and $B$ satisfy are
	\begin{eqnarray}
	\sum_{v_1v_2 \eta_1\eta_2\eta_4} T_{\eta_1 \eta_2 \eta_3 \eta_4} \tilde{B}_{\eta_2v_1v_1'} \tilde{B}^{\ast}_{\eta_4v_2v_2'} A_{\eta_1 v_1v_2} &=& \lambda_A A_{\eta_3 v_1'v_2'},\\
	\sum_{v_1v_{1'}\eta_1\eta_2\eta_3} T_{\eta_1 \eta_2 \eta_3 \eta_4} A_{\eta_1 v_1v_2} A_{\eta_3 v_1'v_2'} B_{\eta_2v_1v_1'} &=& \lambda_B B_{\eta_4v_2v_2'},
	\label{eq6}
	\end{eqnarray} 	
	with $\lambda_A$ and $\lambda_B$ the eigenvalues. The third sub-figure is the QR decomposition $B_{\eta vv'} = \sum_{v''} \tilde{B}_{\eta v'v''} R_{v''v}$, which ensures that $A$ and $B$ converges to the non-trivial fixed-points \cite{S. J. Ran16}. This define in fact a recursive dynamics: after randomly initializing $A$ and $B$, the fixed-point can be reached by recursively solving the above equations.
	
	Note that there is a redundant gauge freedom on the shared bonds between $A$ and $B$. In order to remove it, we define the ring tensor $R$ from $A$ and $B$ [Fig. \ref{fig1} (d)] as
	\begin{equation}
	R_{\eta_1\eta_2 \eta_3\eta_4} = \sum_{v_1v_2v_1'v_2'} A_{\eta_1 v_1v_2}A^{\ast}_{\eta_3 v_1'v_2'} B_{\eta_2 v_1v_1'} B^{\ast}_{\eta_4 v_2v_2'}.
	\end{equation}
	In the tensor ring decomposition, $A$ represents the ``ground state'' MPS of the TN at horizontal direction ; $B$ is the time-MPS at the vertical direction\cite{S. J. Ran17, M. B. Hastings07, E Tirrito16}. Both are also known as the boundary states of the TN \cite{BMPS1, BMPS2, BMPS3, RanTNcrit}. Meanwhile, the ring tensor $R$ is actually an approximation of the environment of one tensor $T$, i.e., the tensor after contracting all the TN without $T$. Thus, the contraction $Z = \sum_{\eta_1\eta_2 \eta_3\eta_4} T_{\eta_1\eta_2 \eta_3\eta_4} R_{\eta_1\eta_2 \eta_3\eta_4}$ gives approximately the whole TN contraction, and it is maximized at the fixed-point.

	\section{Ring degeneracy and global symmetry}
	One may expect there is only one ring tensor for a TN since it represents the contraction of TN. However, when the local tensor $T$ of TN have a symmetry, the symmetry may induce a degeneracy on ring tensors. The ring degeneracy $\mathcal{D}$ is then defined by the number of ring tensors that give the same partition function $Z(R) := Tr(RT)$. It can be checked by the fidelity $F$ of two ring tensors $R$ and $R'$ [Fig. \ref{fig1} (e)] as
	\begin{equation}
	F(R, R') = |\sum_{\eta_1\eta_2 \eta_3\eta_4} R_{\eta_1\eta_2 \eta_3\eta_4} R_{\eta_1\eta_2 \eta_3\eta_4}^{\prime \ast}| / \sqrt{|R| |R'|}.
	\end{equation}
	
\begin{figure}
		\includegraphics[angle=0,width=1\linewidth]{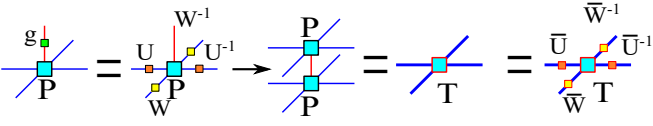}
		\caption{(Color online) The illustration of the global symmetry of a TNS and its inner-product TN [see Eqs. (\ref{eq4}) and (\ref{eq5})].}
		\label{fig3}
	\end{figure}

	Suppose an injective TNS $|\psi\rangle$ satisfies a global symmetry $G$, which requires the tensor $P$ to satisfy the following condition \cite{D. Perez-Garcia08} (Fig. \ref{fig3})
	\begin{eqnarray}
	&&\sum_{s'}V_{ss'}^{[g]} P_{s',\alpha_1 \alpha_2 \alpha_3 \alpha_4} = \nonumber \\ 
	&&\sum_{\alpha_1' \alpha_2' \alpha_3' \alpha_4'} U^{[g]}_{\alpha_1 \alpha_1'} (U^{[g]})^{-1}_{\alpha_3 \alpha_3'} W^{[g]}_{\alpha_2 \alpha_2'} (W^{[g]})^{-1}_{\alpha_4 \alpha_4'} P_{s,\alpha_1' \alpha_2' \alpha_3' \alpha_4'}.
	\label{eq4}
	\end{eqnarray}
	Here $g$ is a group element of $G$ and $V^{[g]}$ is a representation of $g$; $U$ and $W$ are the projective representation of the group respected to $g$ \cite{X. G. Wen02, D. Perez-Garcia08, X. Chen11-2}. In this case, the tensor $T$ in the inner product TN $\langle \psi | \psi \rangle$ possesses the corresponding symmetry $T = \mathcal{G}(T)$ (Fig. \ref{fig3}) that reads
	\begin{eqnarray}
	\mathcal{G}(T) \equiv \sum_{\eta_1' \eta_2' \eta_3' \eta_4'} \bar{U}^{[g]}_{\eta_1 \eta_1'} (\bar{U}^{[g]})^{-1}_{\eta_3 \eta_3'} \bar{W}^{[g]}_{\eta_2 \eta_2'} (\bar{W}^{[g]})^{-1}_{\eta_4 \eta_4'} T_{\eta_1' \eta_2' \eta_3' \eta_4'}.
	\label{eq5}
	\end{eqnarray}
	
	with $\bar{U}^{[g]} = U^{[g]} \otimes U^{[g]\ast}$ and $\bar{W}^{[g]} = W^{[g]} \otimes W^{[g]\ast}$. For tensor $T' = \mathcal{G}(T)$, we can always define a ring tensor $R' = \mathcal{G}^{-1}(R)$ even when $T \neq \mathcal{G}(T)$. However, when $T$ satisfies the symmetry condition as $T = \mathcal{G}(T)$, $R'$ is also a ring tensor of $T$. Thus, the ring degeneracy emerges when $F[R, \mathcal{G}(R)] \neq 1$.
	
	\section{Ring degeneracy and symmetry breaking in Ising model}
	We first apply our method to the 2D statistical Ising model on square lattice, where the TN satisfies the $Z_2$ symmetry. This model was investigated by Gu \textit{et al} \cite{ZC.Gu09} as the very first example that inspired the (trivial) symmetry-protected topological orders. The interaction of this model is described by $H = \sum_{\langle i,j \rangle} \eta_i \eta_j$, where $\eta_i$ represents the Ising spin on the $i$-th site, and the summation runs over all nearest-neighbor pairs of spins. The partition function $Z = Tr(e^{-\beta H})$ can be written as TN, where we have $T_{\eta_1\eta_2 \eta_3\eta_4} = e^{-\beta (\eta_1 \eta_2 + \eta_2 \eta_3 + \eta_3 \eta_4 + \eta_4 \eta_1)}$. This Hamiltonian is invariant under a global $Z_2$ transformation, hence the tensor $T$ are also invariant under $Z_2$ transformation. When the temperature $\text{T}$ is higher than the critical temperature $\text{T}_c = \frac{2}{\ln(\sqrt{2}+1)} \approx 2.26919$, the system is in a disordered state; when $\text{T} < \text{T}_c$, there exist two degenerate ordered states that one can be transform into another by applying $Z_2$ spin flip transformation on it, and the system reaches either of it by spontaneously breaking the symmetry.
	
\begin{figure}
		\includegraphics[angle=0,width=1\linewidth]{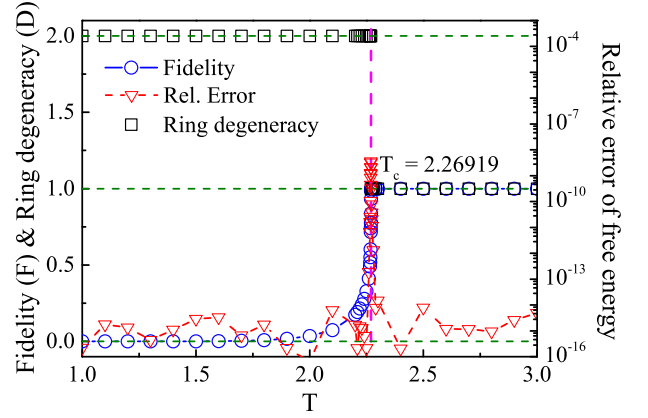}
		\caption{(Color online) The fidelity, relative error of energy and ring degeneracy in the 2D statistical Ising model. For the temperature $\text{T} > \text{T}_c$ (with $\text{T}_c$ the critical temperature), there is only one fixed-point. When $\text{T} \leq \text{T}_c$, the overlap rapidly vanishes $O(10^{-5})$, indicating the existence of two degenerate fixed-points that are orthogonal to each other.The relative errors of the free energy (compared with the analytical solution) are also shown, which is about $O(10^{-9})$ at the critical temperature and soon decays to $O(10^{-15})$ away from $\text{T}_c$.}
		\label{fig2}	
	\end{figure}

	Fig. \ref{fig2} shows the results at different temperatures (with bond dimension cut-off $\chi = 40$). For $\text{T} > \text{T}_c$, $\mathcal{D} = 1$. This is because $R = \mathcal{G}(R)$ and there is only one fixed-point representing the high-temperature disordered phase even when the TN satisfies the symmetry. For $\text{T} < \text{T}_c$, the symmetry of the ring tensor is broken and we obtain $\mathcal{D} = 2$, i.e., the two ring tensors give the same $Z$ and are orthogonal to each other with $F(R, R') \simeq 0$. In the symmetry breaking phase, one ring tensor can be transformed into another by performing a $Z_2$ transformation on it, reflecting two degenerate ground states as the boundary states of the TN.
	
	\section{Ring degeneracy in Z$_N$ topological systems}
	The ground state of spin-1 Heisenberg chain is in the well-known Haldane phase with non-trivial topological orders \cite{F. D. M. Haldane83-1, F. D. M. Haldane83-2}. The Haldane gap is $\Delta_{H} \simeq 0.4105J$ \cite{S. R. White93}. The Hamiltonian (in a magnetic field) reads $H = J\sum_{\langle i,j \rangle}(\textbf{S}_i\textbf{S}_j) + h\sum_{i}S^z_i$. By Trotter-Suzuki decomposition, the imaginary-time evolution of this model can be represented by a 2D TN as tensor product density operator(TPDO)\cite{S. J. Ran17, ODTNS, NCD, FTPEPS, Variant_FTPEPS, FTTNR, FTTNR2} (see for instance \cite{TEBD}). With trotter step $\tau = 0.01$ and bond dimension cut-off $\chi = 100$, the result shows ring degeneracy precisely matches the phase diagram, where we have $\mathcal{D} = 2$ for $h < h_c$ (Haldane phase) and $\mathcal{D} = 1$ for $h > h_c$ with $h_c = 0.4126J \approx \Delta_{H}$.
	
\begin{figure}
		\includegraphics[angle=0,width=1\linewidth]{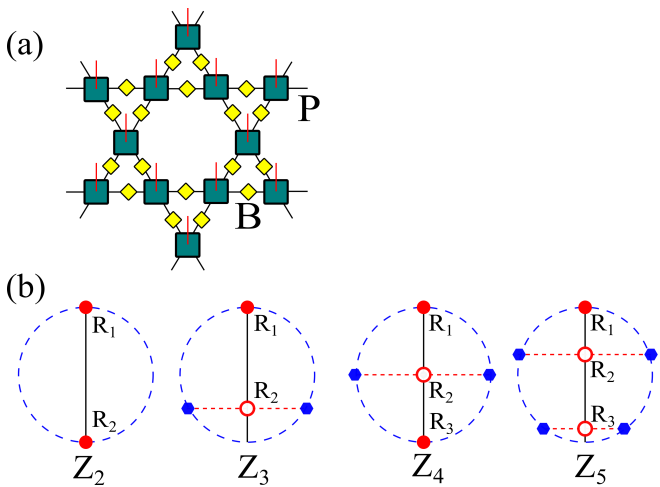}
		\caption{(Color online) (a) Graphic representation of TNS representation of NNRVB state on kagom\'e lattice. (b) An intuitive picture explaining the relation between the ring degeneracy and $Z_N$ orders, by taking $Z_2$, $Z_3$, $Z_4$ and $Z_5$ as examples. The black line represents the real space and the blue dash circle represents the complex space. The blue dots represent the fixed-point solutions (ring tensors) in the complex space, and the red dots show the projections in the real space by combining two conjugate solutions.}
		\label{fig4}
	\end{figure}

	The nearest-neighbor resonating valence bond (NNRVB) state on Kagom\'e lattice is a quantum spin liquid state with intrinsic $Z_2$ topological order \cite{P. W. Anderson73, X. G. Wen91, R. Moessner01, G. Misguich02}. Its TN representation is formed by the infinite copies of tensors $P$ and $B$ (see in Fig. \ref{fig4} (a)), whose non-zero elements are \cite{F. Verstraete06, N. Schuch12}
	\begin{eqnarray}
	\begin{array}{c}
	P_{0,0000} = P_{1,2111} = P_{1,1211} = P_{1,1121} = P_{1,1112} = 1\\
	B_{00} = B_{12} = 1, B_{21} = -1
	\end{array}
	\end{eqnarray}
	We calculated the TRD of the TN $\langle \psi|\psi \rangle$ with $\chi = 40$ and obtain $\mathcal{D}=2$. The fidelity between the two degenerate ring tensors is $F \sim 10^{-9}$.
	
	The $Z_N$ string-net states \cite{M. Levin05,M. Levin06, Z. C. Gu09,O. Buerschaper09} possess intrinsic $Z_N$ topological orders \cite{X. Chen11,X.Chen10}. On a square lattice, the TNS of a $Z_N$ string-net state can be defined by the tensor as
	\begin{eqnarray}
	T_{\alpha \beta \gamma \delta}&=&
	\left\{
	\begin{array}{lll}
	1, \ \ (\alpha + \beta + \gamma + \delta) \mod N= 0 \\
	0, \ \ otherwise
	\end{array}
	\right.
	\end{eqnarray}
	We applied our method on these states with $\chi = 40$ and find that the ring degeneracy $\mathcal{D}$ satisfies
	\begin{eqnarray}
	\mathcal{D}&=&
	\left\{
	\begin{array}{lll}
	(N + 2)/ 2, \ \ \text{N is even} \\
	(N + 1 )/2, \ \ \text{N is odd}
	\end{array}
	\right.
	\label{eq-relation}
	\end{eqnarray}
	
	To better understand the even-odd pattern in Eq. (\ref{eq-relation}), we give an intuitive picture (Fig. \ref{fig4} (b)) and explain it by the representation theory of the $Z_N$ group. For $Z_N$ group, the group elements can be represented as $\{ \mathbb{I}, g, g^2, \cdots, g^{N-1}\}$. From the representation theory, all irreducible representations of $Z_N$ are one dimensional and can be denoted by $g^k = \exp(ik\theta)$ with $N\theta = 0(\text{mod } 2\pi)$. Hence for $N > 2$, the non-trivial representation of $Z_N$ group should be complex. However, when applying TRD in the complex space, we meet with a convergence problem. Our results show that there exist several fixed-points, where the fidelity between each two can be any values between 0 and 1. The reason might be that the fixed-points ``drifts'' due to the gauge degrees brought by a complex phase factor. Thus, we restrain ourselves in the real space, and the fidelity takes only $0$ or $1$. In this case, the gauge degrees of freedom are fixed due to the uniqueness of the dominant eigenvectors of the two eigenvalue problems.
	
	In the even cases, there always exist two real transformation operators: identity $g_0 = \mathbb{I}$ and inversion $g_{N/2} = -\mathbb{I}$, which give two real ring tensors noted as $R$ and $\mathcal{G}_{N/2}(R)$. By projecting on the real space, a real solution can be defined by the superposition of a complex ring tensor $\mathcal{G}_k(R)$ and its conjugate $\mathcal{G}_{-k}(R)$. In this way, $(N-2)$ complex tensors will give us $(N-2)/2$ projected ring tensors. In total, there will be $(N-2)/2+2=(N+2)/2$ real fixed-points. When $N$ is odd, there is only one real operator as the identity $\mathbb{I}$, and $N-1$ complex transformations will give $(N-1)/2$ projected ring tensors. Thus the degeneracy of the ring tensors will be $(N-1)/2+1 = (N+1)/2$ in total. 
	
	Though this even-odd pattern of $\mathcal{D}$ makes $Z_{2N}$ state and $Z_{2N+1}$ state share the same $\mathcal{D}$, we can still identify these two cases by examining the partition function $Z(R)$. For $Z_N$ case, the real ring tensor $R_{real}$ (red dots in Fig. \ref{fig4}) gives the partition function $Z(R_{real}) = N$; for projected ring tensor $R_{proj}$ (red circles) we have $Z(R_{proj}) =  N/2$. Thus, $Z_{2N}$ and $Z_{2N+1}$ states can be distinguished as the following: by checking all $\mathcal{D}$ ring tensors and the partition function $Z(R)$ given by them. If we can find two ring tensors that can gave $Z(R) = 2N$ it is a $Z_{2N}$ state, else if there is only one ring tensor that gave $Z(R) = 2N+1$, then it is a $Z_{2N+1}$ state
	
	\section{Robustness under pure noises}
\begin{figure}
		\includegraphics[angle=0,width=1\linewidth]{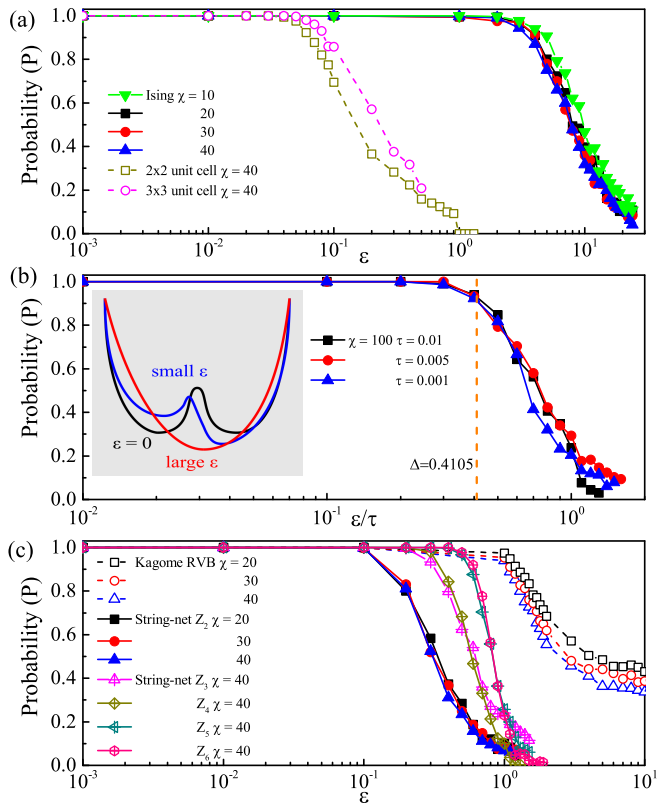}
		\caption{(Color online) (a) The probability $P$ of finding two degenerate fixed-points for the 2D Ising model at the temperature $\text{T} = 1$ with different randomness strength. We take the bond dimension $\chi=10 \sim 40$, and break the translation invariance by selecting the unit cells of sizes $1\times 1$, $2\times 2$, and  $3\times 3$. (b) The probability $P$ for the TPDO of spin-1 Haldane chain, here $\chi = 100$ and we choose three different trotter steps as $\tau = 0.01, 0.005$, and $0.001$ and shows there is an uniform probability under the normalized perturbation parameter as $\epsilon/\tau$. In the inset of (b) we give an intuitive picture of the robustness of RD. For $\epsilon = 0$ there exist two degenerate fixed-points. When a small $\epsilon$ is turned on, it will break the degeneracy but two fixed-points still survive until the $\epsilon$ is sufficiently large. (c) The probability $P$ for the Kagom\'e NNRVB and $Z_N$ ($N = 2, 3, 4, 5$) string-net states with $\chi = 20 \sim 40$.}
		\label{fig5}
	\end{figure}

	To investigate the effect of noises, we add a perturbation term $\epsilon T_p$ to the TN, i.e., $\tilde{T} = T_0 + \epsilon T_p$, with $T_p$ a tensor that all components are chosen randomly with a Gaussian distribution with centered $0$ and standard deviation $1$, and $\epsilon$ a constant to control the strength of the noise. This perturbation term would break the symmetry of local tensors in TNS. Our results show that even though the random term breaks the symmetry of TNS, the fixed-points of the TRD (if exist) remain robust. This robustness can be understood by an intuitive picture shown in the inset of Fig. \ref{fig5} (b). Taking $Z_2$ case as an example, the two degenerate ring tensors $R_1$ and $R_2$ give the same partition function $Z(R_1) = Z(R_2)$. After adding a small noise, the two fixed-points of Eq. (\ref{eq6}) still survive, even though the partition functions is perturbed [$Z(R_1) \neq Z(R_2)$ up to the strength of the noise]. Hence, the RD is robust under noise as long as the fixed-points are still stable contractors of the recursive process in the TRD. 
	
	In the following, we randomly choose about 50 pairs of $A$ and $B$ as the initial guesses to compute the fixed-points. Different initial guesses may be within the attraction domain of different fixed-points. We then check the ring tensors by calculating the fidelity between each two of the fixed-points, and obtain the ring degeneracy as the number of orthogonal ring tensors. To characterize the stability, we define the probability as $P = N_{\mathcal{D} = 2} / N_{tot}$, where $N_{tot}$ is the total number of different random terms $T_p$ we added to $T_0$, and $N_{\mathcal{D} = 2}$ is the number of those terms with which the expected fixed-points are successfully found.
	
	The results of the 2D Ising model [Fig. \ref{fig5} (a)] show that the two fixed-points remain stable (with $P \simeq 1$) for $\epsilon < 0.1$. For $\epsilon > 0.1$, the probability $P$ drops rapidly, and finally decays to zero where the fixed-points are totally destroyed by the noise. Note that TRD applies to the TN that is translation invariance. The TN is formed by the copies of $\tilde{T} = T_0 + \epsilon T_p$, meaning the random terms for different tensors are the same. To weaken such a translation invariance, we increase the unit cell, so that the random terms are transitionally invariant for $L \times L$ tensor clusters. Inside the cluster, the random terms added to different tensors are independent to each other. Our results show that the stability persists for $L=1$, 2 and 3.
	
	For the spin-1 Haldane chain, we use $\epsilon/\tau$ to properly define the strength of the noise, considering the perturbation is added directly to the evolution operator $e^{-\tau \hat{H}}$. Taking different Trotter steps $\tau = 0.01, 0.005$ and $0.001$, a significant drop occurs universally around $\epsilon/\tau \approx 0.4$, which is consistent with the Haldane gap [Fig. \ref{fig5} (b)]. As we know the boundary state of this TN (i.e., the ground state) is in the Haldane phase. The consistency between the Haldane gap and the noise tolerance suggests that the the gap of the boundary protects the RD degeneracy from the noise.
	
	For the nearest-neighbor RVB state on kagom\'e lattice, we try more than $2000$ different $T_p$ as the noise and calculate the probability $P(\mathcal{D} = 2)$ with $\chi = 20, 30, 40$ [Fig. \ref{fig5} (c)]. By increasing the randomness strength $\epsilon$ from $10^{-7}$ to $10$, $P( \mathcal{D} = 2)$ is still almost $1$ when $\epsilon = 1$. It decays to about $0.35$ after $\epsilon \geq 10$. The probability of finding two fixed-points on $Z_2$ string-net state starts to decay at $\epsilon = 0.1$ from $P(\mathcal{D} = 2) \simeq 1$ to $P( \mathcal{D} = 2) \leq 0.05$ at $\epsilon = 1.2$. For the $Z_N$ string-net states for $N = 3, 4, 5, 6$, the robustness of the ring degeneracy in shown in Fig. \ref{fig5} (c). The bond dimension is fixed as $\chi = 40$. From the probability $P$, the ring degeneracy remains robust for different $N$ up to a random strength of $O(10^{-1})$. 	
	
	\section{Conclusions}
	We propose a noise-tolerant detection for the $Z_N$ topological orders of quantum many-body states by utilizing the TN representation. This quantity, dubbed as the ring degeneracy, is defined by the degeneracy of the fixed-point solutions of the self-consistent equations that encode the TN contraction. The RD and the symmetry of the ring tensor $R$ reveal non-trivial properties of the system described by the TN. For the 2D Ising model, RD indicates the two degenerate states in low-temperature ordered phase. These states are reflected by the degenerate ground states in the boundary theory of TN. For $Z_N$ topological systems, the RD detects the specific topological orders by the symmetry in accordance to the topological order. It is interesting to notice that, when the TN is an inner product of a 2D quantum state, RD can detect the topological order of the state, and when the TN is partition function of a classical system, RD is detecting the symmetry breaking phase. It suggest a connection between 2D quantum states and partition function of a 2D classical systems.  
	
	Different from the existing quantities such as entanglement spectrum, RD is defined as the number of stable contractors of the self-consistent eigenvalue equations; our data shows it can survive under pure noises up to certain strength. In the Spin-$1$ Haldane chain model, the strength is consistent to the Haldane gap, which suggests the strength of robustness is comparable to the gap of the boundary theory. When the noise breaks the symmetry of TNS, topological ground state degeneracy will be lifted, and topological Renyi entropy will not be observed. However, those lifted states are still the stable fixed-points of the given recursive process. Such a property could be used to investigate the states in the non-symmetrical vicinity of topologically ordered states; it provides a robust detection for the topological properties even when symmetry is slightly broken.  Our work provides a simple and robust detection for the topological orders, and reveals the stability of many-body topology from the perspective of recursive dynamics.
	
	\section{Acknowledgment}
	X. Chen and G. Su are supported in part by the NSFC (Grant No. 11834014), the National Key R\&D Program of China (Grant No. 2018YFA0305800), and the Strategic Priority Research Program of CAS (Grant Nos. XDB28000000, XBD07010100). S. Yang is supported in part by NSFC (Grant N0. 11804181), the National Key R\&D Program of China (Grant No. 2018YFA0306504) and the Research Fund Program of the State Key Laboratory of Low-Dimensional Quantum Physics (Grant No. ZZ201803). M.L. acknowledges the Spanish Ministry MINECO (National Plan 15 Grant: FISICATEAMO No. FIS2016-79508-P, SEVERO OCHOA No. SEV-2015-0522, FPI), European Social Fund, Fundació Cellex, Generalitat de Catalunya (AGAUR Grant No. 2017 SGR 1341 and CERCA/Program), ERC AdG OSYRIS, EU FETPRO QUIC, and the National Science Centre, Poland-Symfonia Grant No. 2016/20/W/ST4/00314. S.J.R. acknowledges Fundaci\'o Catalunya - La Pedrera $\cdot$ Ignacio Cirac Program Chair, Beijing Natural Science Foundation (No. 1192005  and Z180013), and Foundation of Beijing Education Committees under Grants No. KZ201810028043.

\end{document}